\documentclass[11pt]{article}
\usepackage{graphicx}
\usepackage{epsfig}

\newcommand{\BABARPubYear}    {01}

\newcommand{\BABARConfNumber} {09}
\newcommand{\SLACPubNumber} {8980}

\input pubboard/babarsym

\long\def\inst#1{\par\nobreak\kern 4pt\nobreak
    {\it #1}\par\vskip 10pt plus 3pt minus 3pt}

\begin{document}
\pagestyle{empty}


\begin{flushright}
\babar-CONF-\BABARPubYear/\BABARConfNumber \\
SLAC-PUB-\SLACPubNumber \\
September, 2001 \\
\end{flushright}

\par\vskip 5cm

\begin{center}
\Large \bf 
Search for Direct $CP$ Violation \\in Quasi-Two-Body Charmless $B$ Decays
\end{center}
\bigskip

\begin{center}
\large The \babar\ Collaboration\\
\mbox{ }\\
\today
\end{center}
\bigskip \bigskip

\begin{center}
\large \bf Abstract
\end{center}
We have searched for direct $CP$ violation in quasi-two-body 
charmless $B$ decays observed in 
a sample of about 45 million $B$ mesons collected with the 
\babar\ detector at the PEP-II collider.
We measure the following charge asymmetries in decay:
${\cal A}_{CP} ( B^{\pm}\to\eta^\prime K^{\pm} ) = -0.11\pm 0.11\pm 0.02$, \linebreak
${\cal A}_{CP} ( B^{\pm}\to\omega \pi^{\pm} ) = -0.01^{+0.29}_{-0.31}\pm 0.03$,
${\cal A}_{CP} ( B^{\pm}\to\phi K^{\pm} ) = -0.05\pm 0.20\pm 0.03$,\linebreak
${\cal A}_{CP} ( B^{\pm}\to\phi K^{*\pm} ) = -0.43^{+0.36}_{-0.30}\pm 0.06$, and
${\cal A}_{CP} ( B^{0}/\bar B^{0}\to\phi K^{*0}/\bar K^{*0} ) = 
0.00\pm 0.27\pm 0.03$.

\vfill
\begin{center}
Submitted to the 9$^{th}$ International Symposium 
on
Heavy Flavor Physics, \\
9/10---9/13/2001, Pasadena, CA, USA
\end{center}

\vspace{1.0cm}
\begin{center}
{\em Stanford Linear Accelerator Center, Stanford University, 
Stanford, CA 94309} \\ \vspace{0.1cm}\hrule\vspace{0.1cm}
Work supported in part by Department of Energy contract DE-AC03-76SF00515.
\end{center}

\newpage
\pagestyle{plain}

\begin{center}
\small

The \babar\ Collaboration,
\bigskip

B.~Aubert,
D.~Boutigny,
J.-M.~Gaillard,
A.~Hicheur,
Y.~Karyotakis,
J.~P.~Lees,
P.~Robbe,
V.~Tisserand
\inst{Laboratoire de Physique des Particules, F-74941 Annecy-le-Vieux, France }
A.~Palano,
A.~Pompili
\inst{Universit\`a di Bari, Dipartimento di Fisica and INFN, I-70126 Bari, Italy }
% ...

\end{center}\newpage

\section{Introduction}
\label{sec:Introduction}
The phenomenon of $CP$ violation has played an important
role in understanding fundamental physics since its
initial discovery in the $K$ meson system in 1964 \cite{Cronin}.
Recently, a significant $CP$ violating asymmetry 
has been observed in the $B$ meson system~\cite{babarcp}. 
Both effects may be accounted for by a non-zero phase in the
mixing of two neutral mesons ($K^0-\bar{K}^0$ 
or $B^0-\bar{B}^0$). There is a different type of $CP$ violation 
due to interference among decay amplitudes which differ in both weak 
and strong phases. This ``direct" $CP$ violation has been 
observed recently in kaon decays~\cite{directkaon}. 
While $CP$ violation effects are small in the kaon system 
they are anticipated to be larger in $B$ decays \cite{Bander}.
Direct $CP$ violation would be evident in an asymmetry of $B$ 
decay rates:
\begin{eqnarray}
{\cal A}_{CP} = \frac{\Gamma(\bar B\rightarrow\bar f)-\Gamma(B\rightarrow f)}
              {\Gamma(\bar B\rightarrow\bar f)+\Gamma(B\rightarrow f)} \ .
\label{eq:acpdecay}
\end{eqnarray}

Rare $B$ meson decays are particularly interesting in searches for
direct $CP$ violation because they have significant penguin amplitudes.
In the Standard Model substantial $CP$ violation in $B$ decays could 
arise from interference of penguin ($P$) and tree ($T$) amplitudes 
\cite{Bander}:
\begin{eqnarray}
{\cal A}_{CP} = \frac
{2~|P|~|T|~\sin\Delta\phi~\sin\Delta\delta}
{|P|^2 + |T|^2 + 2~|P|~|T|~\cos\Delta\phi~\cos\Delta\delta}
 \ ,
\label{eq:acpphase2}
\end{eqnarray}
where $\Delta\phi$ and $\Delta\delta$ are the differences
in weak and strong phases.
The weak phase difference, $\Delta\phi$, between the $b\to u$ tree 
and $b\to s$ (or $b\to d$) penguin amplitudes 
is $\gamma$ (or $\gamma+\beta$), as in the case of the
decays $B^\pm\rightarrow\eta^\prime K^\pm$ 
(or $B^\pm\rightarrow\omega\pi^\pm$).
Thus, ${\cal A}_{CP}$ is sensitive to the CKM angles 
$\gamma$~and~$\alpha = \pi-(\gamma+\beta)$,
where $\gamma = {\rm arg}(V^*_{ub})$ 
and $\beta = {\rm arg}(V_{td})$ 
in the usual phase convention~\cite{Wolfenstein, Kobayashi}.
However, there is large uncertainty in the strong phases, 
which weakens any quantitative relationship to the weak 
phase angles.

Even more interesting is the scenario of direct $CP$ violation
in the pure penguin modes, such as $B\to\phi{K}^{(*)}$. In the
Standard Model, the expected ${\cal A}_{CP}$ is negligible. However, 
new particles in loops, such as charged Higgs or SUSY particles,
would provide additional amplitudes with different phases. Depending 
on the model parameters, ${\cal A}_{CP}$ may be as large as $30\%$ 
with new physics~\cite{newphys}. Complementary searches for new physics
would involve measurements of the time-dependent asymmetries in $B$ decays
to $CP$ eigenstates, such as $\phi K^0_{S(L)}$ and 
$\eta^\prime K^0_{S(L)}$. Comparison of the value of $\sin 2\beta$ 
obtained from these modes with that from charmonium modes
can probe for new physics participating in penguin loops. 
In these measurements, direct $CP$ violation in the decay becomes 
highly relevant.

A search for direct $CP$ violation in $B$ meson decays to 
$\pi K$, $\eta^\prime K$, and $\omega\pi$ 
was performed previously by the CLEO experiment \cite{cleo}.
In this paper we improve the precision of the measurements
and extend the search for direct $CP$ violation to new modes
with data from the $\babar$ experiment.
We present measurements of the charge asymmetry in the 
quasi-two-body charmless $B$ 
decays~[\ref{ref:phipaper},~\ref{ref:etaprpaper}]:
$B^{\pm}\to\eta^\prime K^{\pm}$, 
$B^{\pm}\to\omega \pi^{\pm}$, 
$B^{\pm}\to\phi K^{\pm}$,
$B^{\pm}\to\phi K^{*\pm}$, and
$B^{0}/\bar B^{0}\to\phi K^{*0}/\bar K^{*0}$.
We choose modes where the $B$ flavor is tagged 
by its charge, except for the $\phi K^{*0}/\bar K^{*0}$ 
final state where the flavor is tagged by the charge of the kaon from 
the $K^{*0}\to K^+\pi^-$ decay.
A measurement from \babar\ of the $\pi K$ charge asymmetry may be found
elsewhere \cite{kpipaper}.

\section{Detector and Data}
\label{sec:Data}
The data were collected with the \babar\ detector~\cite{babar}
at the PEP-II asymmetric $e^+e^-$ collider~\cite{pep}
located at the Stanford Linear Accelerator Center.
The results presented in this paper are based on data taken
in the 1999--2000 run. An integrated
luminosity of 20.7~fb$^{-1}$ was recorded corresponding to 
22.7 million $B\overline{B}$ pairs at the $\Upsilon (4S)$ resonance
(``on-resonance'') and 2.6~fb$^{-1}$ about 40~MeV below
this energy (``off-resonance''). 

The asymmetric beam configuration in the laboratory frame
provides a boost to the $\Upsilon(4S)$
increasing the momentum range of the $B$-meson decay products
up to 4.3~GeV/$c$.
Charged particles are detected and their momenta are measured
by a combination of a silicon vertex tracker (SVT) consisting 
of five double-sided layers and a 40-layer central drift chamber 
(DCH), both operating in a 1.5~T solenoidal magnetic field. 
With the SVT, a position resolution of about 40~$\mu$m is 
achieved for the highest momentum charged particles near the 
interaction point, allowing the precise determination of 
decay vertices.
The tracking system covers 92\% of the solid angle
in the center-of-mass system (CM).
The track finding efficiency is, on average, (98$\pm$1)\% for momenta
above 0.2~GeV/$c$ and polar angle greater than 0.5~rad. 
Photons are detected by a CsI electromagnetic calorimeter (EMC), which
provides excellent angular and energy resolution with high efficiency for 
energies above 20~MeV~\cite{babar}.

Charged particle identification is provided by the average 
energy loss ($dE/dx$) in the tracking devices and
by a unique, internally reflecting ring imaging 
Cherenkov detector (DIRC) covering the central region. 
A Cherenkov angle $K$--$\pi$ separation of better than 4$\sigma$ is 
achieved for tracks below 3~GeV/$c$ momentum, decreasing to 
2.5$\sigma$ at the highest momenta in our final states. 
Electrons are identified with the use
of the tracking system and the EMC.

\section{Event Selection}
\label{sec:Selection}
All the selection requirements are identical to those used in 
the branching fraction 
measurements~[\ref{ref:phipaper},~\ref{ref:etaprpaper}].
Hadronic events are selected based on track multiplicity and 
event topology. We fully reconstruct $B$ meson 
candidates from their charged and neutral 
decay products, where we recover the intermediate states
$\eta^\prime\rightarrow\eta\pi^+\pi^-$ ($\eta^\prime_{\eta\pi\pi}$)
or $\rho^0\gamma$ ($\eta^\prime_{\rho\gamma}$),
$\omega\rightarrow\pi^+\pi^-\pi^0$,
$\phi\rightarrow K^+K^-$,
$K^{*+}\rightarrow K^0\pi^+$ ($K^{*+}_{K^0}$)
or $K^+\pi^0$ ($K^{*+}_{K^+}$), 
$K^{*0}\rightarrow K^+\pi^-$,
$\rho^{0}\rightarrow \pi^+\pi^-$,
$\pi^0\rightarrow \gamma\gamma$,
$\eta\rightarrow \gamma\gamma$, and
$K^0\rightarrow K^0_S\rightarrow\pi^+\pi^-$. 
Candidate charged tracks are required to originate 
from the interaction point, and to have at least 12 DCH hits 
and a minimum transverse momentum of 0.1~GeV/$c$. 
Looser criteria are applied to tracks forming $K^0_S$ candidates
to allow for displaced decay vertices.
Kaon tracks are distinguished from pion and proton tracks via a
likelihood ratio that includes, for momenta below 0.7~GeV/$c$, 
$dE/dx$ information from the SVT and DCH, and, for higher
momenta, the Cherenkov angle and number of photons
as measured by the DIRC. 

We combine pairs of tracks with opposite charge from a common
vertex to form $K^0_S$, $\phi$, $K^{*0}$, and $\rho^0$ candidates.
We further combine a pair of charged tracks with a $\pi^0$ or $\eta$ candidate
to select $\omega$ or $\eta^\prime_{\eta\pi\pi}$ candidates.
The selection of $K^0_S$ candidates is based on the invariant two-pion mass 
($|M_{\pi\pi} - m_{K^0}|<$ 12~MeV/$c^2$), the angle $\alpha$ between 
the reconstructed flight and momentum directions ($\cos\alpha >$ 0.995),
and the measured lifetime significance ($\tau/\sigma_\tau >$ 3).

We reconstruct $\pi^0$ ($\eta$) mesons as pairs of photons
with a minimum energy deposition of 30~MeV (100~MeV). 
The typical width of the reconstructed $\pi^0$ mass is 7~MeV/$c^2$.
A $\pm$15~MeV/$c^2$ interval is applied to select $\pi^0$ candidates. 
We combine a $\rho^0$ candidate with a photon of energy above
$200$~MeV to obtain an $\eta^\prime_{\rho\gamma}$ candidate.

We select $\phi$, $\omega$, $\eta^\prime$, and $\eta$ candidates 
with the following requirements on the invariant masses of
their final states (in MeV/$c^2$): 
$990 < m(K^+K^-) < 1050$, 
$735 < m(\pi^+\pi^-\pi^0) < 830$, 
$930 < m(\eta\pi^+\pi^-) <990$, 
$900 < m(\rho\gamma) <1000$, and
$490 < m(\gamma\gamma) < 600$.
The natural widths of the $K^*$ and $\rho$ dominate the resolution 
in the invariant mass spectrum. We require the invariant $\rho$
mass to be between 500~MeV/$c^2$ and 995~MeV/$c^2$. 
For $K^{*}$ candidates the $K\pi$ invariant mass interval 
is either $\pm 100$ or $150$~MeV/$c^2$ \cite{phipaper}.

The helicity angle $\theta_H$ of a $\phi$, $K^*$, or $\omega$
is defined as the angle between the direction of one
of the two daughters, or the normal to the $\omega$ decay plane,
and the parent $B$ direction in the resonance rest frame. 
To suppress combinatorial background we restrict the 
$K^{*+}\rightarrow K^+\pi^0$ helicity angle 
($\cos\theta_H > -0.5$). This effectively 
requires the $\pi^0$ momentum to be above 0.35~GeV/$c$.

We identify $B$ meson candidates kinematically
using two nearly independent 
variables \cite{babar},
the energy-substituted mass
$m_{\rm{ES}} =$ 
$\sqrt{ (s/2 + \mathbf{p}_i \cdot \mathbf{p}_B)^2 / E_i^2 - 
\mathbf{p}_B^{\,2} }$ and
$\Delta E = (E_i E_B - \mathbf{p}_i 
\cdot \mathbf{p}_B - s/2)/\sqrt{s}$,
where $\sqrt{s}$ is the total $e^+e^-$ CM energy.
The initial-state 
four-momentum $(E_i,\mathbf{p}_i)$
derived from the beam kinematics and the four-momentum
$(E_B,\mathbf{p}_B)$ of the reconstructed $B$ candidate 
are all defined in the laboratory.
An alternative to $m_{\rm{ES}}$ is the energy constrained
mass $m_{\rm{EC}}$, which is obtained from the kinematic
fit of the measured candidate four momentum in the 
$\Upsilon(4S)$ frame with the constraint of the $B$ energy
to the beam energy. Both $m_{\rm{ES}}$ and $m_{\rm{EC}}$
provide almost identical background separation, while
$m_{\rm{EC}}$ is less correlated 
to $\Delta E$ than is $m_{\rm{ES}}$.
For signal events $\Delta E$ peaks at zero and
$m_{\rm{ES}}$ and $m_{\rm{EC}}$ at the $B$ mass. 

Monte Carlo (MC) simulation \cite{geant} demonstrates that contamination
from other $B$ decays is negligible.
However, charmless hadronic modes suffer from large backgrounds due to random 
combinations of tracks produced in the quark-antiquark ($q\bar{q}$) 
continuum.  This background is distinguished by its jet structure as 
compared to the spherical decay of the $\Upsilon$.
To reject continuum
background we make use of the angle $\theta_T$ between the thrust axes
of the $B$ candidate and the rest of the tracks and neutral clusters in
the event, calculated in the center-of-mass frame.  The distribution of
$\cos{\theta_T}$ is sharply peaked near 
$\pm1$ for combinations drawn from jet-like $q\bar q$ pairs, and nearly
uniform for the isotropic $B$ meson decays.
Thus we require $|\cos\theta_T| < 0.9$ (0.8 for $\phi K^{*+}$). 
We also construct a Fisher discriminant \cite{CLEO-fisher}
which combines eleven variables: the angles of the B
momentum vector and the B two-body decay axis with respect 
to the beam axis in the $\Upsilon(4S)$ frame,
and a nine bin representation of the energy flow 
about the $B$ decay axis.

\section{Maximum Likelihood Fit}
\label{sec:Fit}
We use an extended unbinned maximum likelihood (ML) fit to extract
signal yields and charge asymmetries simultaneously.
The extended likelihood for a sample of $N$ events is
\begin{equation}
{\cal L} = \exp\left(-\sum_{i=1}^{M} \sum_{k=1}^2 n_{ik}\right)\, \prod_{j=1}^N 
\left(\sum_{i=1}^M \sum_{k=1}^2 n_{ik}\, 
{\cal P}_{ik}(\vec{x}_j;\vec{\alpha})\right) ,
\label{eq:likel}
\end{equation}
where ${\cal P}_{ik}(\vec{x}_j;\vec{\alpha})$ describes the probability
for candidate event $j$ to belong to category $i$
and flavor state $k$, 
based on its measured variables $\vec{x}_j$, and fixed parameters
$\vec{\alpha}$ that describe the expected distributions of these
variables in each of the $M$ categories. 
This probability is non-zero only for the right final 
state flavor ($k = 1$ for $\bar B\rightarrow\bar f$ and 
$k = 2$ for $B\rightarrow f$).
In the simplest case, the probabilities are summed over two
categories ($M=2$), signal and background.
The decays with the charged primary daughter $h^\pm$ ($K^\pm$ or $\pi^\pm$)
are fit simultaneously with two signal and two corresponding 
background categories ($M=4$). These are: 
$B^\pm\rightarrow\eta^\prime h^\pm$, $\omega h^\pm$, and $\phi h^\pm$.
We rewrite the event yields $n_{ik}$ in each category in terms of 
the asymmetry ${\cal A}_i$ and the total event yield $n_{i}$:
$n_{i1} = n_{i}\times(1 + {\cal A}_i)/2$ and
$n_{i2} = n_{i}\times(1 - {\cal A}_i)/2$.
The event yields $n_i$ and asymmetries ${\cal A}_i$
in each category are obtained by maximizing ${\cal L}$ \cite{minuit}.
Statistical errors correspond to unit changes in the quantity 
$\chi^2 = -2\ln{{\cal L}}$ around its minimum value. 
The significance of non-zero asymmetry is defined 
by the square root of the change in $\chi^2$ when 
constraining the asymmetry to zero in the likelihood fit.
The 90\% C.L. limits correspond to a change in $\chi^2$ of 2.69.
When more than one channel is measured for the same
primary $B$ decay, the channels are combined
with $\chi^2$ distributions.

The probability ${\cal P}_{ik}(\vec{x}_j;\vec{\alpha})$ 
for a given event $j$ is the product of independent probability
density functions (PDFs) in each of the fit input variables $\vec{x}_j$.
These are $\Delta E$, $m_{\rm{ES}}$ or $m_{\rm{EC}}$,
invariant masses of intermediate resonances 
($\eta^\prime$, $\omega$, $\phi$, $K^*$, and $\eta$), 
Fisher discriminant, and the $\phi$ and $\omega$ helicity angles for 
pseudoscalar-vector decays. 
For the simultaneous fit to the decays with the charged
primary daughter $h^\pm$ we include normalized residuals 
derived from the difference between measured and expected 
DIRC Cherenkov angles for the $h^\pm$.
Additional separation between the two final states is provided by 
$\Delta E$, where the separation depends on the momentum of the 
charged primary daughter in the laboratory and it is about 
45 MeV on average.

The fixed parameters $\vec{\alpha}$ describing the PDFs are extracted 
from signal and background distributions from MC 
simulation, on-resonance $\Delta E$--$m_{\rm{ES}}$ sidebands, and
off-resonance data. 
The MC resolutions are adjusted by comparisons of data and simulation 
in abundant calibration channels with similar kinematics and topology,
such as $B\rightarrow D\pi, D\rho$ with $D\rightarrow K\pi, K\pi\pi$.
The simulation reproduces the event-shape variable distributions
found in data.
The Cherenkov angle residual parameterizations are 
determined from samples of $D^0\rightarrow K^-\pi^+$
originating from $D^*$ decays.  

For the parameterization of the PDFs for $\Delta E$,
$m_{\rm{ES}}$ or $m_{\rm{EC}}$, and resonance masses we employ
Gaussian and Breit-Wigner functions to describe the
signal distributions. 
For the background we use low-degree polynomials or,
in the case of $m_{\rm{ES}}$ or $m_{\rm{EC}}$, 
an empirical phase-space function~\cite{argus}.
The background parameterizations for resonance masses
also include a resonant component to 
account for resonance production in the continuum.
In the $B$ decays into vector-vector states, the 
$\cos\theta_H$ distribution is the result of an 
{\it a priori\/} unknown superposition of transverse 
and longitudinal polarizations, and thus it is not used
for background suppression in the fit.
For pseudoscalar-vector $B$ decay modes, angular momentum
conservation results in a $\cos^2\theta_H$ distribution 
for signal. 
The background shape is again separated into contributions 
from combinatorics and from real mesons, both
fit by nearly constant low-degree polynomials.  
The Cherenkov angle residual PDFs are Gaussian
for both the pion and kaon distributions.
The Fisher discriminant is described by an asymmetric Gaussian 
for both signal and background.

\section{Results}
\label{sec:Results}
\begin{figure}[htbp]
\begin{center}
\centerline{
\psfig{file=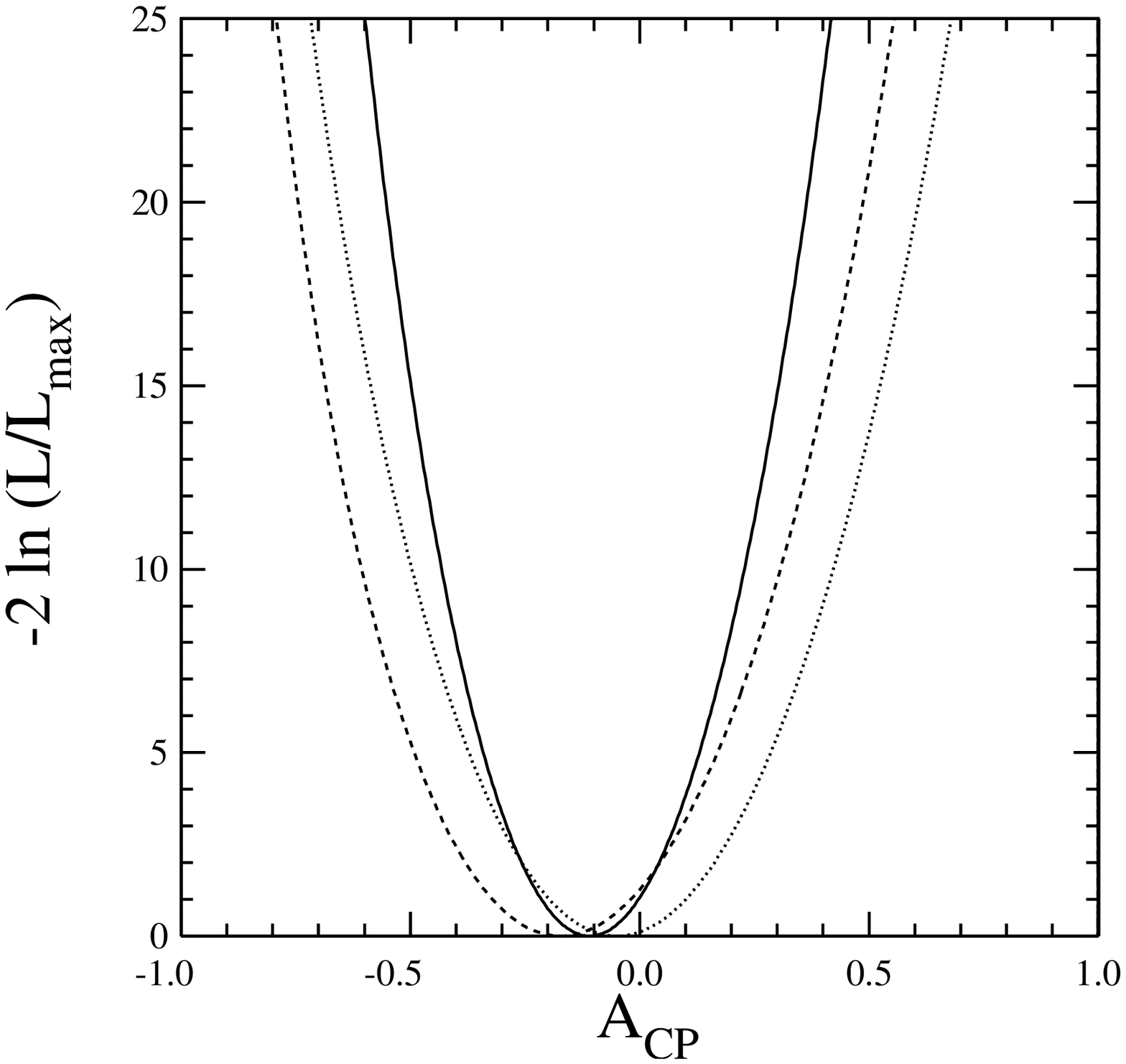,width=2.8in}
\psfig{file=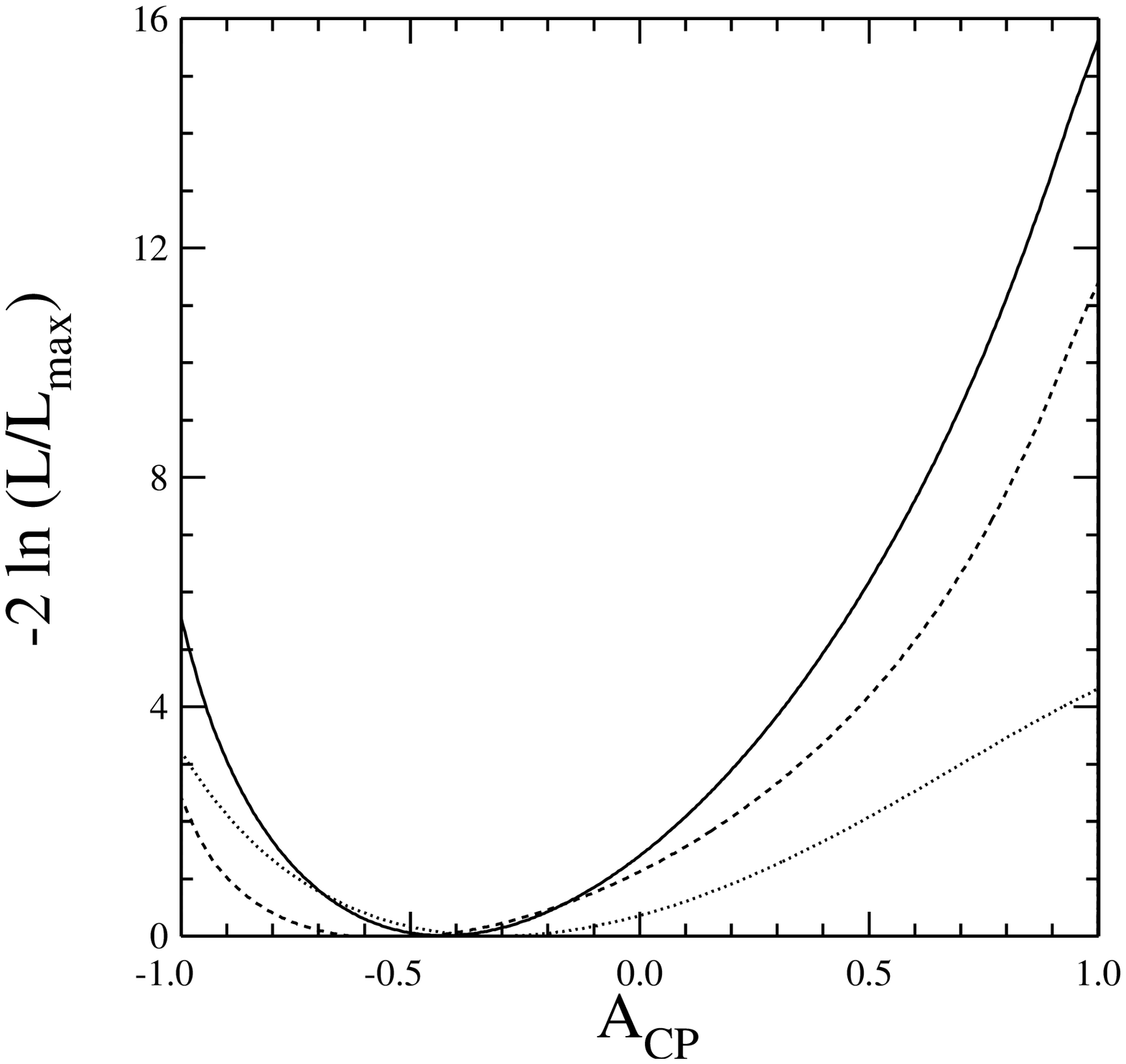,width=2.8in}
}
\centerline{
\psfig{file=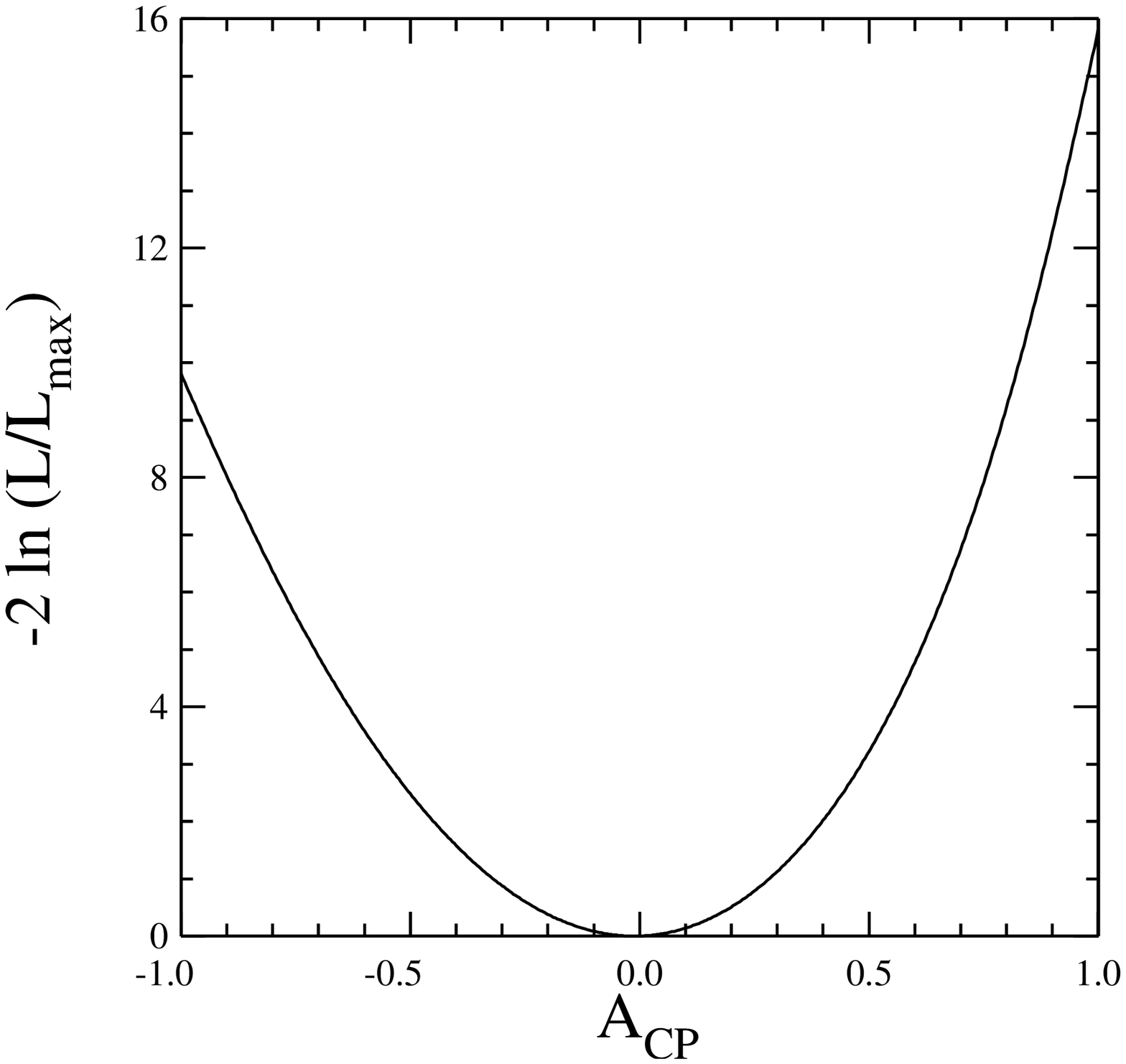,width=1.8in}
\psfig{file=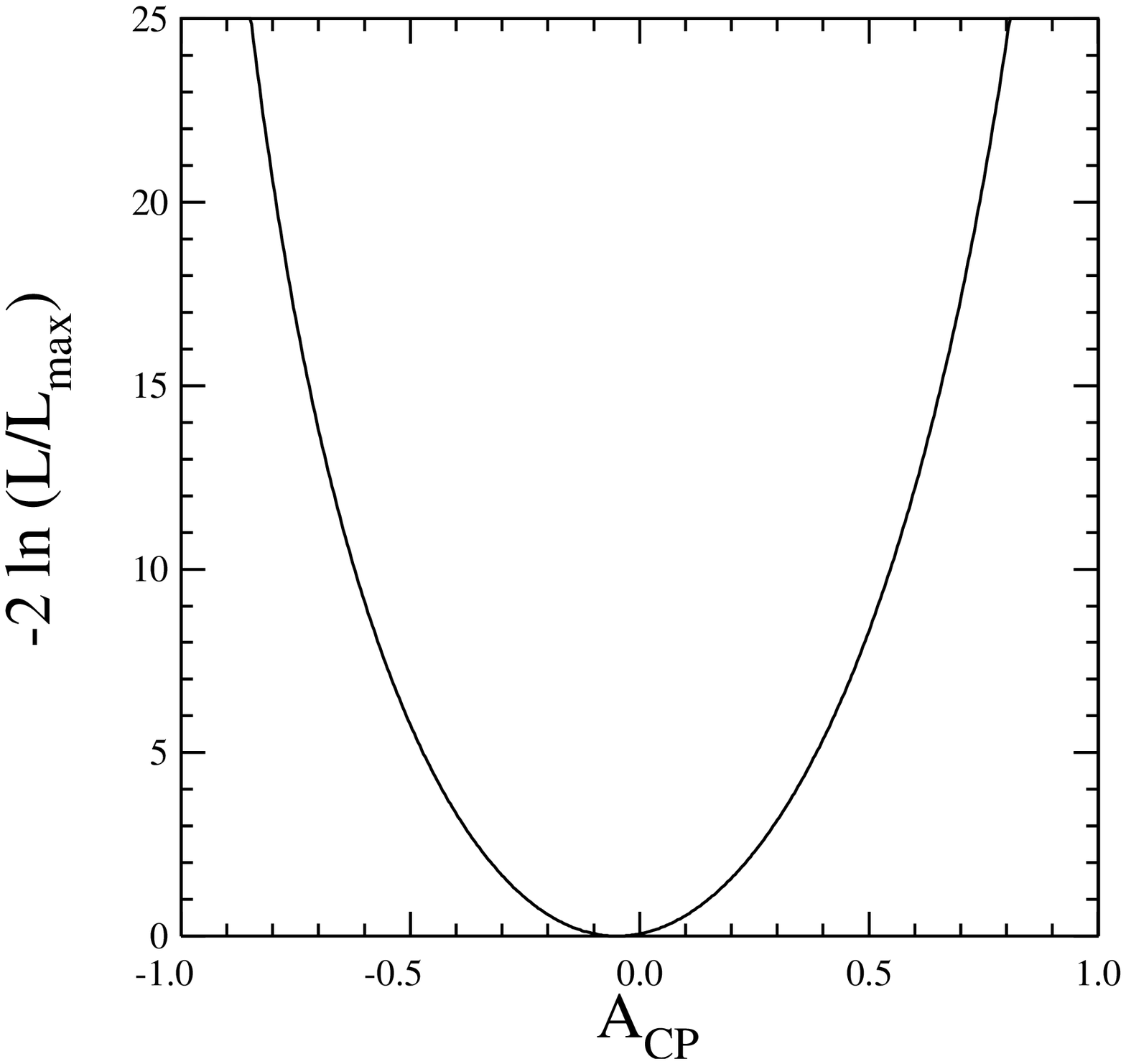,width=1.8in}
\psfig{file=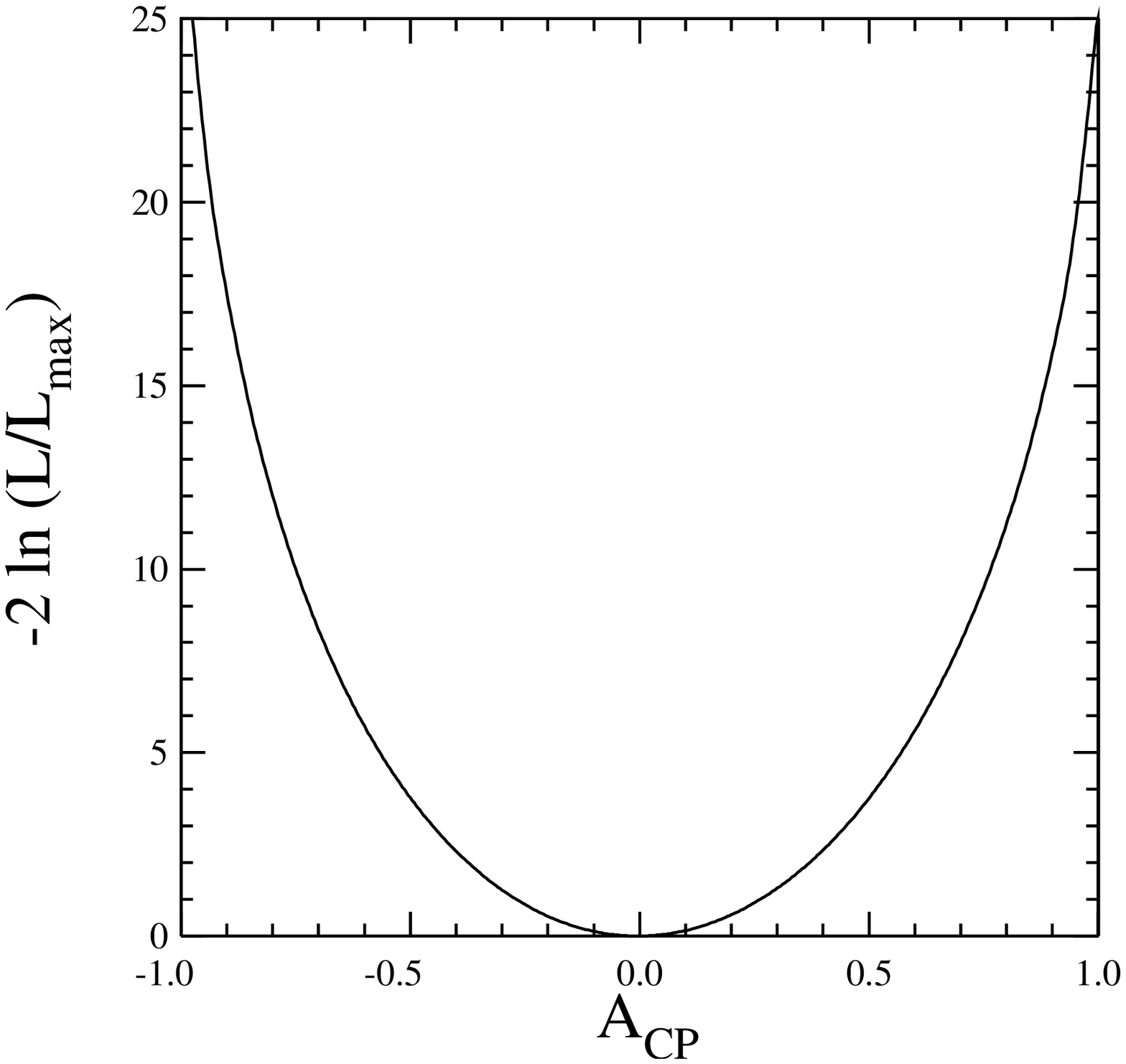,width=1.8in}
}
\caption
{ Distribution of $\chi^2$ for the charge asymmetries 
${\cal A}_{CP}$ in the physically allowed range.
Top plots: two secondary channels (dashed and dotted lines)
are combined to produce a final result (solid line);
left plot: $\etapr K^\pm$ 
with  $\etapr_{\eta\pi\pi} K^\pm$ (dashed)
and $\etapr_{\rho\gamma} K^\pm$ (dotted);
right plot: $\phi K^{*\pm}$ 
with $\phi K_{K^0}^{*\pm}$ (dashed)
and $\phi K_{K^+}^{*\pm}$ (dotted).
Bottom plots, from left to right:
$\omega\pi^\pm$, $\phi K^{\pm}$, and $\phi K^{*0}/\bar{K}^{*0}$.
}
\label{fig:acpchi}
\end{center}
\end{figure}

\begin{figure}[htbp]
\begin{center}
\centerline{\epsfig{figure=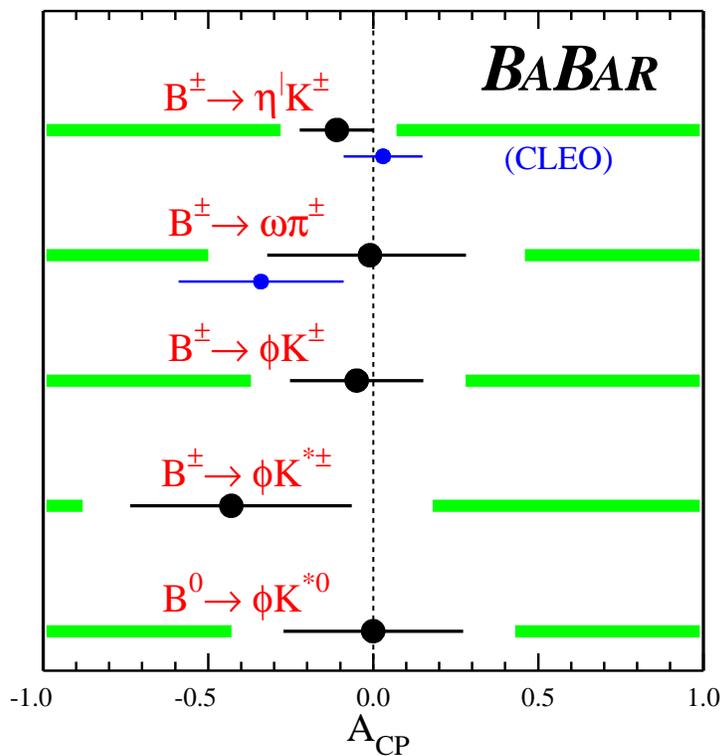,width=4.0in}}
\caption
{ Results of the direct $CP$ violation search
in the $B$ decays into final states $\eta^\prime K^{\pm}$, 
$\omega \pi^{\pm}$, $\phi K^{\pm}$, $\phi K^{*\pm}$, and
$\phi K^{*0}/\bar K^{*0}$.
Points with error bars represent experimental 
measurements of ${\cal A}_{CP}$. Solid thick 
lines delimit the 90\% C.L. intervals. For the modes
$\eta^\prime K^{\pm}$ and $\omega \pi^{\pm}$ 
smaller points with error bars show results
of the CLEO experiment \cite{cleo}.
}
\label{fig:acp_visual}
\end{center}
\end{figure}

The results of our ML fit analyses are summarized in 
Table~\ref{tab:results}. The signal yields along with
branching fraction results have been reported earlier 
\cite{phipaper, etaprpaper}.
In all cases we find significant signal event yields, 
and hence proceed with asymmetry measurements. 
The dependence of the $\chi^2$ on 
${\cal A}_{CP}$ for each decay mode and sub-channel
is shown in Fig.~\ref{fig:acpchi} and asymmetry
measurements are summarized in Fig.~\ref{fig:acp_visual}.
We see no significant asymmetries and set 
90\% C.L. intervals.


\begin{table}[h!]
\caption
{Results of the ML fits:
number of signal events ($n_{\rm sig}$), their charge
asymmetry (${\cal A}_{CP}$), asymmetry 90\% C.L. limits and
significance ($S_{\cal A}$).
All results include systematic errors.
}
\label{tab:results}
\begin{center}
\begin{tabular}{lcccc}
\hline\hline
\vspace{-3mm}&&&\\
Mode & $n_{\rm sig}$ & ${\cal A}_{CP}$ & 90\% C.L. & $S_{\cal A}$
($\sigma$) \cr
\vspace{-3mm}&&&\\
\hline
\vspace{-3mm}&&&\\
$\etapr K^\pm$ & -- 
 & $-0.11\pm 0.11\pm 0.02$ & [--0.28;+0.07] & 1.0  \cr
\vspace{-3mm}&&&\\
~~~$\etapr_{\eta\pi\pi} K^\pm$ & $49.5^{+8.1}_{-7.3}\pm 1.5$ 
 & $-0.17\pm 0.15 \pm 0.01$ & -- & 1.1  \cr
\vspace{-3mm}&&&\\
~~~$\etapr_{\rho\gamma} K^\pm$ & $87.6^{+13.4}_{-12.5}\pm 3.7$ 
 & $-0.05\pm 0.15\pm 0.03$ & -- & 0.3  \cr
\vspace{-3mm}&&&\\
\hline
\vspace{-3mm}&&&\\
$\omega\pi^\pm$ & $27.6^{+8.8}_{-7.7}\pm 1.9$ 
 & $-0.01^{+0.29}_{-0.31}\pm 0.03$ & [--0.50;+0.46] & 0.0  \cr
\vspace{-3mm}&&&\\
\hline
\vspace{-3mm}&&&\\
$\phi K^\pm$ & $31.4^{+6.7}_{-5.9}\pm 2.3$ 
 & $-0.05\pm 0.20\pm 0.03$ & [--0.37;+0.28] & 0.2  \cr
\vspace{-3mm}&&&\\
\hline
\vspace{-3mm}&&&\\
$\phi K^{*\pm}$ & -- 
 & $-0.43^{+0.36}_{-0.30}\pm 0.06$ & [--0.88;+0.18] & 1.2  \cr
\vspace{-3mm}&&&\\
~~~$\phi K_{K^0}^{*\pm}$ & $4.4^{+2.7}_{-2.0}\pm 0.4$ 
 & $-0.55^{+0.51}_{-0.35}\pm 0.05$ & -- & 1.1  \cr
\vspace{-3mm}&&&\\
~~~$\phi K_{K^+}^{*\pm}$ & $7.1^{+4.3}_{-3.4}\pm 1.2$ 
 & $-0.31^{+0.54~+0.10}_{-0.43~-0.06}$ & -- & 0.6  \cr
\vspace{-3mm}&&&\\
\hline
\vspace{-3mm}&&&\\
$\phi K^{*0}$ & $20.8^{+5.9}_{-5.1}\pm 1.3$ 
 & $0.00\pm 0.27\pm 0.03$ & [--0.43;+0.43] & 0.0  \cr
\vspace{-3mm}&&&\\
\hline\hline
\end{tabular}
\end{center}
\end{table}

Most of the systematic error contributions relevant to
branching fraction analyses cancel 
for the ratio in Eq.~\ref{eq:acpdecay}.
Some level of charge asymmetry bias is inevitable as
neither the $\babar$ detector nor PEP-II is perfectly 
charge symmetric. However these effects are mostly very 
small for the final states considered here. Charge 
biases in the detector and track reconstruction 
have been studied in a sample of more than a billion charged tracks 
in multi-hadron events.  After proton and electron
rejection we find an asymmetry consistent with zero with an uncertainty
of less than 1\% for a wide range of momenta. Taking into account particle
identification requirements, this consistency is still better
than 2\%. The $D^{*\pm}$ control sample of kaon and pion 
tracks is used to estimate systematic uncertainties 
in the asymmetries arising from possible charge biases in the
Cherenkov angle residual, which are found to be less than 1\%.

From these studies we assign a systematic uncertainty 
of 1\% on ${\cal A}_{CP}$ for all the modes with 
a charged primary daughter: 
$B^\pm\rightarrow\eta^\prime h^\pm$, $\omega h^\pm$, and $\phi h^\pm$.
For the modes with a $K^*$ we account for the broader momentum spectrum 
of the charged daughters and particle identification applied to 
the kaon candidates with a 2\% systematic error.
All measured background asymmetries and signal asymmetries in MC 
are consistent with zero within statistical uncertainties.

A different type of uncertainty originates in the ML fit 
from assumptions about the signal and background distributions.
We vary the PDF parameters within their respective uncertainties,
and derive the associated systematic errors in the 
event yield and its asymmetry. Corresponding systematic
errors on asymmetry are found to be 
2\% for $\eta^\prime K^\pm$ and $\phi K^{*0}$,
3\% for $\omega\pi^\pm$ and $\phi K^\pm$,
and 6\% for $\phi K^{*\pm}$, the latter being dominated
by the mode with a $\pi^0$.
These systematic errors are conservatively estimated 
and may be improved with higher statistics.

We combine the correlated (due to selection requirements) 
and uncorrelated (due to PDF variations) systematic errors,
and convolute systematic errors into $\chi^2$ distributions 
in order to obtain results with systematics. We also treat
the correlated and uncorrelated systematic errors separately
when we combine the sub-channels. The uncertainties in the
final results presented in Table~\ref{tab:results} are 
dominated by statistical errors.

\pagebreak

\section{Conclusions}
\label{sec:Conclusions}
We have searched for direct $CP$ violation in quasi-two-body 
charmless $B$ decays observed in $\babar$ data.
The measured charge asymmetries 
of the $B$ decays into final states $\eta^\prime K^{\pm}$, 
$\omega \pi^{\pm}$, $\phi K^{\pm}$, $\phi K^{*\pm}$, and
$\phi K^{*0}/\bar K^{*0}$ are summarized in Table~\ref{tab:results}
and Fig.~\ref{fig:acp_visual}. 
The 90\% C.L. limits rule out a significant
part of the physical ${\cal A}_{CP}$ region.

\section{Acknowledgements}
\label{sec:Acknowledgements}
We are grateful for the 
extraordinary contributions of our \pep2\ colleagues in
achieving the excellent luminosity and machine conditions
that have made this work possible.
The collaborating institutions wish to thank 
SLAC for its support and the kind hospitality extended to them. 
This work is supported by the
US Department of Energy
and National Science Foundation, the
Natural Sciences and Engineering Research Council (Canada),
Institute of High Energy Physics (China), the
Commissariat \`a l'Energie Atomique and
Institut National de Physique Nucl\'eaire et de Physique des Particules
(France), the
Bundesministerium f\"ur Bildung und Forschung
(Germany), the
Istituto Nazionale di Fisica Nucleare (Italy),
the Research Council of Norway, the
Ministry of Science and Technology of the Russian Federation, and the
Particle Physics and Astronomy Research Council (United Kingdom). 
Individuals have received support from the Swiss 
National Science Foundation, the A. P. Sloan Foundation, 
the Research Corporation,
and the Alexander von Humboldt Foundation.

\renewcommand{\baselinestretch}{1}

\end{document}